\documentclass[prb,preprint]{revtex4-1} 

\usepackage{amsmath}  
\usepackage{amsfonts}
\usepackage{graphicx}

\begin{document}

%
%

\title{Bidirectional invisibility in Kramers-Kronig optical media}

\author{S. Longhi}
\email{stefano.longhi@polimi.it}
\affiliation{Dipartimento di Fisica, Politecnico di Milano and Istituto di Fotonica e Nanotecnologie del Consiglio Nazionale delle Ricerche, Piazza L. da Vinci 32, I-20133 Milano, Italy}

\date{\today}

%
%
\begin{abstract}
	A broad class of planar dielectric media with complex permittivity profiles that are fully invisible, for both left and right incidence sides, is introduced. Such optical media are locally isotropic, non-magnetic  and belong to the recently discovered class of Kramers-Kronig media [{\it Nature Photon.} 9, 436 (2015)], i.e. the spatial profiles of the real and imaginary parts of the dielectric permittivity are related each other by a Hilbert transform. The transition from unidirectional to bidirectional invisibility, and the possibility to realize sharp reflection above a cut-off incidence angle, are also discussed.
\end{abstract}

\maketitle

%
%

Wave reflection in optical media that show a sharp change of the refractive index is ubiquitous in optics \cite{r0}.  Several methods have been devised to avoid reflection, such as the use of stratified media, antireflection coatings, graded-index and nanostructured interfaces, to mention a few (see, for instance, \cite{r01,r02}). Since the pioneering work by Kay and Moses \cite{r1}, it is well known that a broad class of dielectric media with specially tailored refractive index profiles are reflectionless and can thus provide omnidirectional antireflection \cite{r2,r3}. However, Kay and Moses optical media are not invisible albeit they do not scatter any wave. Invisibility and cloaking devices are generally considered peculiar to metamaterials, which are designed by transformation optics and conformal mapping methods  \cite{m1,m2,m3,m4}. Even in certain isotropic inhomogeneous dielectric media, i.e. with no features of the magnetic permeability,  cloaking can be realized \cite{m5}. Recent works have considered wave reflection in inhomogeneous media with a complex dielectric permittivity profile $\epsilon(x)$ and showed that they can appear invisible when probed form one side (unidirectional invisibility) \cite{r4,r5,r6,r6bis,r7,r8,r8bis,r9,r10,r11}. As compared to metamaterials, they require appropriate dispersion engineering in space but all
 materials are locally isotropic, non-magnetic and do not rely on negatively refracting media. Important examples of one-way invisible dielectric media with a complex dielectric permittivity profile $\epsilon(x)$ are $\mathcal{PT}$-symmetric complex crystals \cite{r5,r6,r7} and Kramers-Kronig optical potentials \cite{r9,r10,r11}. The latter refer to a rather broad class of inhomogeneous planar dielectric media such that the real and imaginary parts of the permittivity $\epsilon(x)$ are related by Kramers-Kronig relations \cite{r9}, i.e. $\epsilon(x)$ is a holomorphic function of the complex variable $x=x'+ix''$ in a half (upper or lower) complex plane. Such media are always reflectionless when probed from one side \cite{r9}, and they turn out to be also invisible when the so-called 'cancellation condition' is met \cite{r10,r11}. It has been also noticed that some  special and exactly-solvable optical potentials \cite{r9,r10}, such as those synthesized by supersymmetry \cite{r10,r12,r12bis},  can show {\it bidirectional} invisibility, i.e. they are invisible regardless of the incidence side. Since in such media there is optical gain, some instability issues might arise \cite{r10,r12}. However, to date no general conditions are known for a complex optical potentials $\epsilon(x)$ to show bidirectional (rather than simple unidirectional) invisibility.
In this Letter we introduce a rather broad class of Kramers-Kronig optical potentials that show {\it bidirectional} invisibility for both TE- and TM-polarized waves. We also discuss the transition from unidirectional to bidirectional invisibility and highlight the possibility to realize sharp reflection above a cut-off angle.\\
Let us consider the scattering of a monochromatic optical wave at frequency $\omega= k_0 c_0$ across an inhomogeneous isotropic planar dielectric medium in the $(x,y)$ plane; see Fig.1. Let $\epsilon=\epsilon(x)$ be the relative dielectric permittivity profile of the medium, which shows an inhomogeneity localized at around $x=0$, i.e.
\begin{equation}
\epsilon(x)= \epsilon_b +\alpha(x)
\end{equation}
where $\epsilon_b=n_b^2$,  $n_b$ is the refractive index of the substrate and $\alpha(x)$ describes the localized inhomogeneity, with $\alpha(x) \rightarrow 0$ as $x \rightarrow \pm \infty$. If dissipation in the substrate is negligible, $\epsilon_b$ is real. On the other hand, $\alpha(x)$ is taken complex, i.e. in  the inhomogeneous region the medium displays optical gain and/or dissipation. We further assume that the limits $\lim_{L \rightarrow \pm \infty} \int_{0}^{L} \alpha(x) dx$ do exist and are finite, which ensures that far from the inhomogeneous region the scattered states are plane waves \cite{r10}.\\ 
The main result of the present work can be stated as follows:\\ 
{\em Let 
\begin{equation}
\alpha(x)=\beta(x) \exp(i \Theta x)
\end{equation}
with $\beta(x)$ holomorphic in the upper half complex plane ${\rm Im}(x) \geq 0$ with $\beta(x) \rightarrow 0$ as $|x| \rightarrow \infty$, and $\Theta \geq  2 k_0n_b$. Then the medium is bidirectionally invisible for TE-polarized waves. In addition, if $\epsilon(x)=\epsilon_b+\alpha(x)$ does not have zeros in the  the upper half complex plane ${\rm Im}(x) \geq 0$, then the medium is bidirectionally invisible for TM-polarized waves as well.}\\
Note that the optical medium satisfying the above requirement is of Kramers-Kronig type, i.e. the real and imaginary parts of $\alpha(x)$ are related by a Hilbert transform, and that the Fourier spectrum $\hat{\alpha}(k)$ of $\alpha(x)$, $\alpha(x)=\int dk \hat {\alpha }(k) \exp(ikx)$, vanishes for $k<\Theta$.
A rigorous proof of such a result is given below. However, it is worth first providing a simple physical explanation of the bidirectional invisibility. Since the Fourier spectrum of the scattering potential $\alpha(x)$ does not have any component with negative wave number $k$, a forward-propagating wave (left incidence side) can not be back scattered. This is basically the physical reason of the one-way reflectionless property of Kramers-Kronig optical media already discussed in Ref.\cite{r9}; invisibility is further ensured by the 'cancellation condition' $\int_{-\infty}^{\infty} dx \alpha(x)=0$ \cite{r10,r11}, which means that the inhomogeneity does not change the amplitude and phase of the transmitted wave. For a backward-propagating wave (right incidence side), there can not be neither forward-propagating scattered waves since the condition $\Theta > 2 k_0 n_b$ makes the scattered waves {\it evanescent}. 
\begin{figure}[htb]
\centerline{\includegraphics[width=8.4cm]{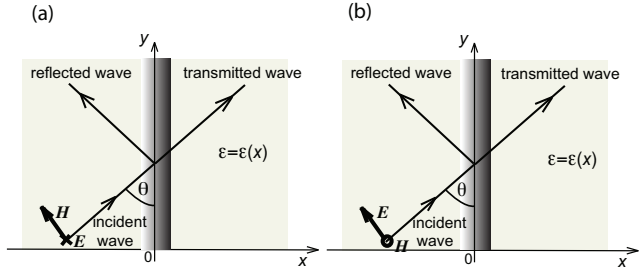}} \caption{ \small
(Color online) Schematic of wave scattering in an inhomogeneous planar dielectric medium for (a) TE-polarized, and (b) TM-polarized waves.}
\end{figure} 

To prove such a result in a rigorous manner, we take the electric and magnetic fields in the form $\mathcal{E}(x,y,z,t)=\mathbf{E}(x,y) \exp(-i \omega t)+c.c.$, $\mathcal{H}(x,y,z,t)=\mathbf{H}(x,y) \exp(-i \omega t)+c.c.$ with invariance along the $z$ direction. Let us first consider the case of a TE-polarized wave ($E_x=E_y=H_z=0$), with $E_z$ satisfying the Helmholtz equation 
$(\partial^2_x+\partial^2_y)E_z+k_0^2 \epsilon E_z=0$; Fig.1(a). After setting $E_z(x,y)=\psi(x) \exp(i k_y y)$, with $0<k_y < k_0 n_b$, the electric field profile $\psi(x)$ satisfies the stationary Schr\"{o}dinger-like wave equation
\begin{equation}
-(d^2 \psi / dx^2)+V(x) \psi=k_x^2 \psi
\end{equation}
with the optical potential given by 
\begin{equation}
V(x)=-k_0^2 \alpha(x)
\end{equation}
and with $k_x=\sqrt{k_0^2 n_b^2-k_y^2}>0$. Note that the incidence angle $\theta$ is given by $\theta= {\rm atan} (k_x/k_y)$, with $\theta \rightarrow 0$ for grazing incidence (Fig.1). With the Ansatz $\psi(x)=w_1(x) \exp(ik_x x)+w_2(x) \exp(-ik_x x)$ and $(d \psi/dx)= ik_x [w_1(x) \exp(ik_x x)-w_2(x) \exp(-ik_xx)]$, the local amplitudes $w_{1,2}(x)$ of forward and backward propagating waves satisfy the exact coupled equations
\begin{eqnarray}
\frac{dw_1}{dx} & = & V(x)/(2ik_x) \left[ w_1 + w_2 \exp(-2ik_x x) \right] \\
\frac{dw_2}{dx} & = & -V(x)/(2ik_x) \left[  w_2 + w_1 \exp(2ik_x x) \right] 
\end{eqnarray}
The transmission $t(k_x)$ and reflection $r^{(l,r)}(k_x)$ coefficients, for left $(l)$ and right $(r)$ incidence sides, are given by
$t(k_x)= w_1(\infty)$,  $r^{(l)}(k_x)=w_2(-\infty)$ with the boundary conditions $w_1(-\infty)=1$, $w_2(\infty)=0$, and $t(k_x)= w_2(-\infty)$,
$r^{(r})(k_x)=w_1(\infty)$ with the boundary conditions $w_2(\infty)=1$, $w_1(-\infty)=0$ (the transmission coefficient is independent of incidence side). To compute the transmission and reflection coefficients, we use the method of spatial complex displacement \cite{r11,r13}, i.e. we introduce the shifted potential $V_1(x)=V(x+i \delta)$ with $\delta>0$ arbitrarily large. 
Then the reflection and transmission coefficients of the original potential $V(x)$ can be retrieved from those $t_1(k_x)$, $r_1^{(l,r)}(k_x)$ of the shifted potential $V_1(x)$ via the simple relations \cite{r11,r13}
\begin{eqnarray}
t(k_x) & = & t_1(k_x) \\
r^{(l)}(k_x) & = & r^{(l)}_{1}(k_x) \exp(-2 \delta k_x) \\
 r^{(r)}(k_x) & = & r^{(r)}_{1}(k_x) \exp(2 \delta k_x).
\end{eqnarray}
Using Eq.(2), the displaced potential $V_1(x)$ can be written as $V_1(x)=F(x) \exp(- \delta \Theta)$ with $F(x)=-k_0^2 \beta(x+ i \delta) \exp(i \Theta x)$. Since $\beta(x)$ is analytic in the ${\rm Im}(x) \geq 0$ half complex plane, $F(x)$ is a limited function for any $\delta>0$, and thus $V_1(x)$ is vanishingly small, at least of order $\eta \equiv \exp(-\Theta \delta)$, in the large $\delta$ limit. Therefore, as $\delta \rightarrow \infty$, for the vanishing potential $V_1(x)$ we can solve the coupled equations (5) and (6), with $V(x)$ replaced by $V_1(x)$, in power series of $\eta$ with appropriate boundary conditions for $w_1(x)$ and $w_2(x)$. At first order (Born) approximation one obtains 
\begin{equation}
t_1(k_x)  =  1+O(\eta) 
\end{equation}
\begin{equation}
r_1^{(l,r)}(k_x)  =  - \eta \left[  \frac{k_0^2}{2ik_x} \int_{-\infty}^{\infty} d \xi \beta( \xi+i \delta ) \exp( \pm 2 i k_x \xi+i \Theta \xi ) \right] + O(\eta^2) \;\;\;\;\;
\end{equation}
where on the right hand side of Eq.(11) the upper (plus) sign applies to $r_1^{(l)}$, whereas the lower (minus) sign applies to $r_1^{(r)}$.
 Note that, since $k_x$ is bounded from above by $k_0 n_b$ and $\Theta>2 k_0 n_b$, the integral on the right hand side in Eq.(11) can be evaluated by the residue theorem closing the contour path
 with a semi-circle of radius $R$ in the ${\rm Im}(\xi) \geq 0$ half complex plane  and using Jordan's lemma. Since $\beta( \xi+ i \delta)$ is holomorphic in the ${\rm Im}(\xi) \geq 0$ half complex plane, using a procedure similar to the one detailed in \cite{r10} and taking the $R \rightarrow \infty$ limit it follows that the integral on the right hand side in Eq.(11) vanishes. 
 Therefore, taking into account that $\Theta > 2 k_x$ and because $\delta$ can be taken arbitrarily large, from Eqs.(7-11) one concludes that $r^{(l)}(k_x)=r^{(r)}(k_x)=0$ and $t(k_x)=1$ for any $k_x>0$.\\
 Let us now consider the case of a TM-polarized incident wave, i.e. let us assume $H_x=H_y=E_z=0$, with $H_z$ satisfying the equation $\epsilon \partial^2_x(H_z / \epsilon) + \partial^2_yH_z+k_0^2 \epsilon H_z=0$; Fig.1(b).  In this case, after setting $H_z(x,y)= \sqrt{\epsilon(x)} \psi(x) \exp(ik_yy)$, the field amplitude $\psi(x)$ satisfies again the stationary Schr\"{o}dinger-like wave equation (3), but with the optical potential $V(x)$ given by $V(x) =-k_0^2 \alpha_{TM}(x)$ where we have set
 \begin{equation}
 \alpha_{TM}(x)= \alpha(x)-\frac{3}{4k_0^2} \left( \frac{\dot{ \epsilon}}{\epsilon} \right)^2+\frac{1}{2k_0^2} \frac{\ddot \epsilon}{\epsilon}.
 \end{equation}
 In the previous equation, the dot indicates the derivative with respect to $x$. Note that, as compared to the TE-polarized case, the optical potential $V(x)$ is modified by the replacement $\alpha(x) \rightarrow \alpha_{TM}(x)$. If we assume for $\alpha(x)$ the form given by Eq.(2) and, {\em additionally}, that $\epsilon(x)$ does not have zeros in the ${\rm Im}(x) \geq 0$ half complex plane, then it can be readily shown that $\alpha_{TM}(x)$ can be written in the form $\alpha_{TM}(x)= \beta_{TM}(x) \exp( i \Theta x)$ with $\beta_{TM}(x)$ holomorphic in the ${\rm Im}(x) \geq 0$ half complex plane. Therefore, likewise
 for the TE polarization case, it follows that the optical medium is bidirectional invisible for TM-polarized waves as well. 
  
\begin{figure}[htb]
\centerline{\includegraphics[width=8.2cm]{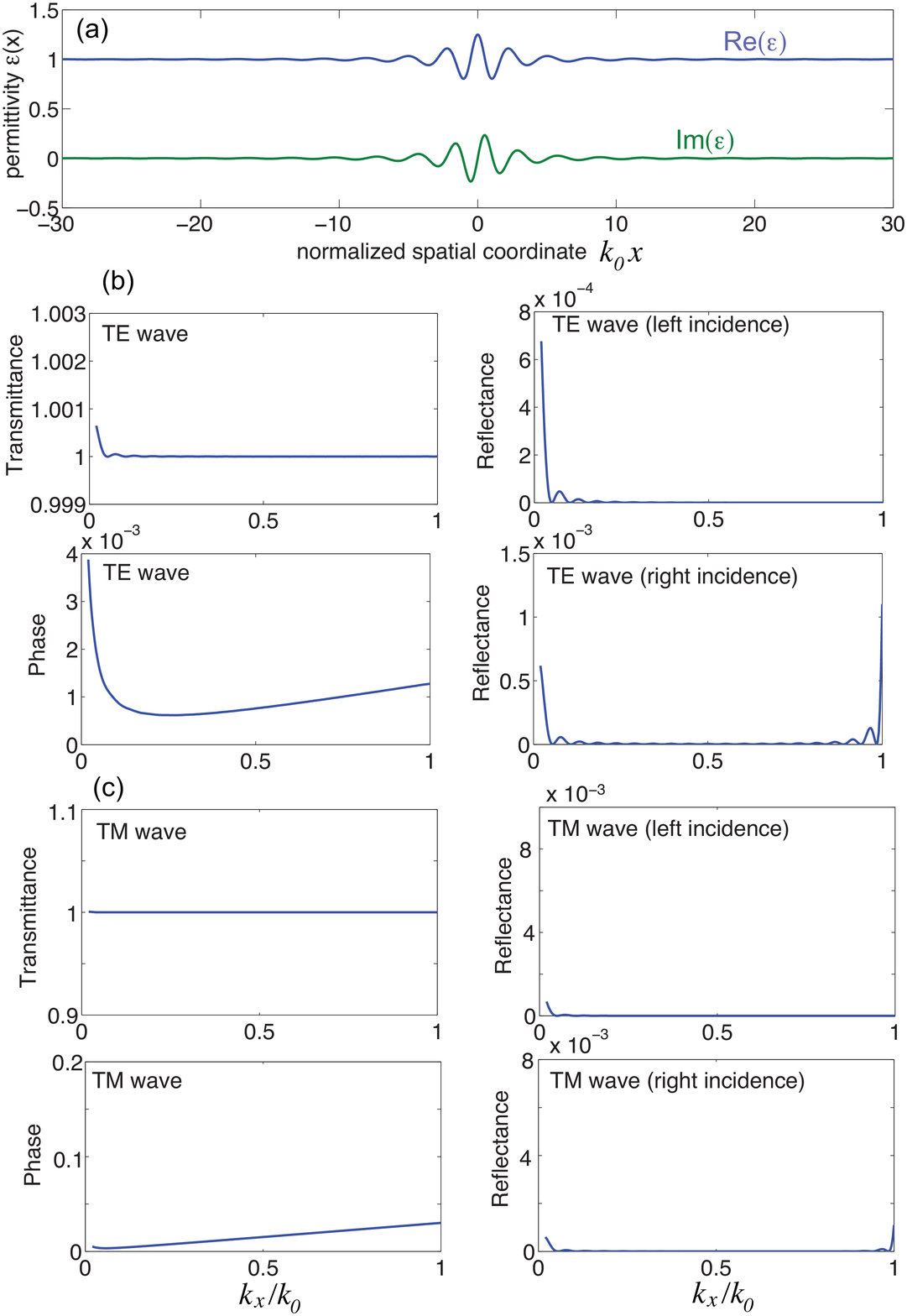}} \caption{ \small
(Color online) Bidirectional invisibility in a Kramers-Kr\"{o}nig planar dielectric medium. (a) Profile of the dielectric permittivity $\epsilon(x)$ defined by Eq.(13). Parameter values are given in the text. (b) Numerically-computed transmittance $|t(k_x)|^2$, reflectances $|r^{(l,r)}(k_x)|^2$ for left and right incidence sides, and phase of the transmission coefficient $t(k_x)$ for a TE-polarized incident wave. (c) Same as (b), but for a TM-polarized wave.}
\end{figure}
 
 As an example, in Fig.2 we show the reflection and transmission coefficients versus $k_x/k_0$, for either TE and TM polarization states, for the dielectric profile 
 \begin{equation}
 \epsilon(x)=n_b^2- A{(x+i \sigma)^{-2}}  \exp(i \Theta x).
 \end{equation}
  The reflection and transmission coefficients have been numerically-computed using a standard transfer matrix method after truncating the potential at $x= \pm L$. Parameter values used in the simulations are $n_b=1$, $\sigma k_0=2$, $\Theta=2k_0$, $Ak_0^2=1$ and $Lk_0=30$. The figure clearly shows bidirectional invisibility for both TE and TM polarized waves.\\
 The previous analysis can be extended to show the transition from unidirectional to bidirectional invisibility in Kramers-Kronig optical potentials. To this aim, let us consider the dielectric permittivity profile $\epsilon(x)=\epsilon_b+ \alpha(x)$ with $\alpha(x)$ defined again by Eq.(2), but we just require now $\Theta > 0$ rather than the more stringent requirement $\Theta \geq  2 k_0 n_b$. Since $\beta(x)$ is holomorphic in the $\rm{Im}(x) \geq 0$ half complex plane, its Fourier spectrum $\hat{\beta}(k)$ vanishes for $k<0$. For the sake of definiteness, let us assume that $\hat{\beta}(k)$ does not vanish in the neighbor of $k=0^+$, and let us consider a TE-polarized wave. 
For $\Theta \rightarrow 0^+$ the medium is invisible for left incidence side whenever $\int_{-\infty}^{\infty} dx \beta(x)=0$, i.e. $\hat{\beta}(k=0)=0$, but rather generally reflection does not vanish for right incidence side \cite{r9}. For $\Theta>0$,  we can again use the method of spatial complex displacement by introduction of the shifted potential $V_1(x)=V(x+i \delta)$, with $\delta>0$ arbitrarily large. Using Eqs.(7-11), one can conclude that $t(k_x)=1$ and $r^{(l)}(k_x)=0$ for any $k_x>0$, i.e. for any incidence angle, whereas $r^{(r)}(k_x)=0$ for $0<k_x< \Theta/2$. This means that, for $0<\Theta< 2 k_0 n_b$, the medium is invisible for left incidence side, whereas for a right incident wave it is only {\it partially} invisible for incidence angle $\theta$ below the cut-off value 
\begin{equation}
\theta_{cut}={\rm atan} \left( \Theta / \sqrt{4 k_0^2n_b^2 -\Theta^2} \right).
 \end{equation}   
 It is also interesting to compute the value of $r^{(r)}(k_x)$ at $k_x= \Theta/2^+$, i.e. at the cut off angle from above. Using Eqs.(9) and (11), one readily obtains
\begin{eqnarray}
r^{(r)}(k_x= \Theta/2^+) & = & \frac{i k_0^2}{\Theta} \int_{-\infty}^{\infty} d\xi \beta(\xi) \exp(-i \xi 0^+) \nonumber \\
&= & \frac{2 \pi i k_0^2}{\Theta}\hat{\beta}(k=0^+).
\end{eqnarray} 
 Hence, whenever the Fourier spectrum $\hat{\beta}(k)$ of $\beta(x)$ does not vanish at $k=0^+$ (for example when $\beta(x)$ shows a first-order pole in the ${\rm Im}(x)<0$ half complex plane), $ r^{(r)}(k_x= \Theta/2^+)$ does not vanish, and thus $r^{(r)}(k_x)$ shows an abrupt change at $k_x=\Theta /2$, from zero at $k_x= \Theta/2^-$ to the value given by Eq.(15) at $k_x= \Theta/2^+$. On the other hand, whenever the Fourier spectrum $\hat{\beta}(k)$ of $\beta(x)$ vanishes at $k=0^+$ (for example when the sum of residues of $\beta(x)$  in the ${\rm Im}(x)<0$ half complex plane vanishes), one has  $ r^{(r)}(k_x= \Theta/2^+)=r^{(r)}(k_x= \Theta/2^-)=0$, i.e. the reflectance for right incidence side is a continuous function when the cut-off angle $\theta_{cut}$ is crossed. From a physical viewpoint, the different behavior of the reflection coefficient near the cut-off angle $\theta_{cut}$ in the two above-mentioned cases stems from the fact that, for an incident wave with momentum $k_x \simeq \Theta /2$, back scattering is provided by the Fourier component of the scattering potential with spatial frequency $2k_x= \Theta$ (elastic scattering). Therefore, a discontinuity of $\hat{\alpha}(k)=\hat{\beta}(k-\Theta)$ at $k= \Theta$ is transferred into an abrupt change of the reflectivity near the incidence angle $\theta \sim \theta_{cut}$. We note that a similar behavior, i.e. continuous or abrupt change of the reflectance depending on the properties of poles of the scattering potential, was found in Ref.\cite{r14} in a different scattering system. As an example, in Fig.3(a) we show the numerically-computed behavior of the reflectance versus $k_x$ for right incidence side in the dielectric permittivity profile defined by Eq.(13) for the same parameter values as in Fig.2, except for $\Theta=k_0$. Note that the medium is invisible only for $k_x<\Theta/2=k_0/2$. Moreover, according to the theoretical analysis the reflectance is a a continuous function at $k_x=\Theta/2$ since $\epsilon(x)$ has a second-order pole at $x=-i \sigma$. Figure 3(b) shows the numerically-computed behavior of the reflectance versus $k_x$ for right incidence side in the dielectric permittivity profile defined by
 \begin{equation}
 \epsilon(x)=n_b^2- A{(x+i \sigma)^{-1}}  \exp(i \Theta x).
 \end{equation}
 with $\Theta=k_0$ as in the previous case. Note that the medium is invisible for $k_x<\Theta/2=k_0/2$, however the reflectance shows an abrupt change (discontinuity) at $k_x=k_0/2$, in agreement with the theoretical predictions [note that $\epsilon(x)$ has a simple pole at $x=-i \sigma$]. For TE-polarized waves, the numerically-computed discontinuity of the reflectance at $k_x=k_0/2$ is in agreement with the theoretical value predicted by Eq.(15).

\begin{figure}[htb]
\centerline{\includegraphics[width=8.4cm]{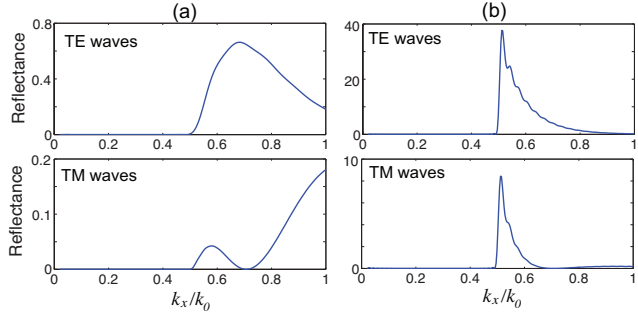}} \caption{ \small
(Color online) Partial invisibility in a Kramers-Kr\"{o}nig planar dielectric medium. (a) Numerically-computed reflectance $|r^{(r)}(k_x)|^2$ for right-side incidence in the permittivity profile defined by Eq.(13) for parameter values $n_b=1$, $\sigma k_0=2$, $\Theta =k_0$ and $Ak_0^2=1$. (b) Same as (a), but for the the permittivity profile defined by Eq.(16) [$n_b=1$, $\sigma k_0=2$, $\Theta= k_0$ and $Ak_0=1$]. In (a) the transition from invisibility to reflection at $k_x= k_0/2$ is continuous, whereas in (b) it is sharp. The reflectances for left incidence side (not shown in the figure) are vanishing, whereas the transmission is unit. In (b) Fabry-Perot like oscillations of reflectance versus $k_x$ can be observed, which arise from the oscillations of the slowly-decaying permittivity profile.}
\end{figure}

\begin{figure}[htb]
\centerline{\includegraphics[width=8.6cm]{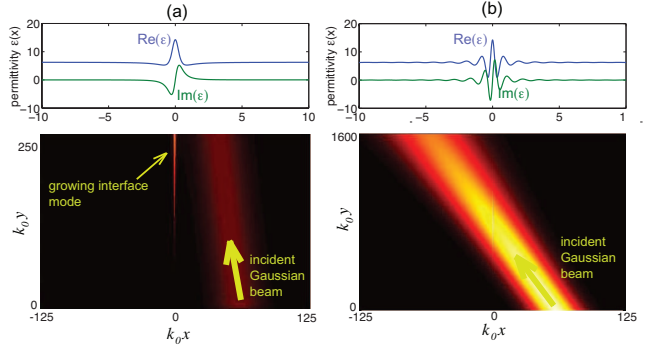}} \caption{ \small
(Color online) Reflection of a TE-polarized Gaussian beam at grazing incidence in the permittivity profile defined by Eq.(13) for (a) $\Theta=0$, and (b) $\Theta= 2 k_0 n_b$. The other parameter values are given in the text. In (a) a secular growth of a localized mode at the interface $x=0$ is observed, whereas the instability is suppressed in (b).}
\end{figure}
As a final comment, it should be noted that in some special Kramers-Kronig potentials, defined by Eq.(2) with $\Theta=0$, the medium turns out to be bidirectionally invisible. An example is provided by the profile defined by Eq.(13), which turns out to be bidirectionally invisible owing to supersymmetry for $\Theta=0$ and for special values of the amplitude $A$, namely $Ak_0^2=n(n+1)$ with $n=1,2,...$ \cite{r10,r15}. However, at grazing incidence such a potential shows an unstable growth of the optical power versus $y$ near the interface $x=0$, which arises from the existence of an exceptional point  in the continuum \cite{r10,r15}. An example of such an instability is shown in Fig.4(a), which depicts the propagation of a TE-polarized broad Gaussian beam across the inhomogeneous region at grazing incidence (incidence angle $\theta \simeq 0.08$ rad) for the permittivity profile defined by Eq.(13) and for parameter values $Ak_0^2=2$, $\sigma k_0=0.5$, $n_b=2.5$ and $\Theta=0$. Note the secular growth of a localized mode near the interface $x=0$. By changing the value of $\Theta$ from $\Theta=0$ to $\Theta=2k_0n_b$, the potential remains bidirectionally invisible according to our general theory. Interestingly, with such a change of $\Theta$ the instability turns out to be suppressed, as shown in Fig.3(b). \par
In conclusion, we have shown that bidirectional invisibility can be observed in a broad class of Kramers-Kronig dielectric and isotropic optical media, and have discussed the transition from unidirectional to bidirectional invisibility. Our results provide new insights into the recently disclosed class of Kramers-Kronig optical media \cite{r9} and should provide guidance in the development
of metamaterials based on judiciously chosen susceptibility profiles.\\
\\
The author acknowledges S.A.R. Horsley for fruitful discussions and suggestions.

\end{document}